\documentstyle[preprint,aps,epsf]{revtex}                  
\global\let\epsfloaded=Y 
\begin{document}
\pagestyle{empty}                                      
\preprint{
\font\fortssbx=cmssbx10 scaled \magstep2
\hbox to \hsize{
\hfill $\raise .5cm\vtop{              
                \hbox{NTUTH-97-05}}$ 
}
}
\draft
\vfill
\title{
Enhanced $b\to sg$ Decay, Inclusive $\eta^\prime$ Production,\\
and the Gluon Anomaly
}

\vfill
\author{Wei-Shu Hou and B. Tseng}
\address{
\rm Department of Physics, National Taiwan University,
Taipei, Taiwan 10764, R.O.C.
}

%
%
\vfill
\maketitle
\begin{abstract}
The experimental hint of large $B\to \eta^\prime + X_s$ is 
linked to the $b\to s$ penguins via the gluon anomaly.
Using running $\alpha_s$ in the $\eta^\prime$-$g$-$g$ coupling,
the standard $b\to sg^*$ penguin alone seems insufficient,
calling for the need of dipole $b\to sg$ at 10\% level from new physics,
which could also resolve the  
${\cal B}_{s.l.}$ and charm counting problems.
The intereference of standard and new physics contributions
may result in {\it direct} CP asymmetries at 10\% level,
which could be observed soon at B Factories.
\end{abstract}
%
%
%
%
\pagestyle{plain}


In this paper we explore the possible connections between 
several fascinating topics in B physics and QCD:
the possibility of enhanced $b\to sg$ decay 
at $10\%$ level\cite{GH,Kagan,KR},
the recent experimental hint \cite{CLEO,Tom,Miller} of large 
inclusive $B\to \eta^\prime + X_s$, 
and the 
gluon anomaly.
It has been suggested \cite{AS} that the standard QCD penguin
could account for inclusive $\eta^\prime$ production 
through the latter. 
We point out that with running $\alpha_s$ in the 
anomaly coupling, however, the Standard Model (SM) alone
may not be sufficient, suggesting the need for
dipole $b\to sg$ transitions from new physics.
If so,
CP violating rate asymmetries between 
$B\to \eta^\prime + X_s$ and $\bar B \to \eta^\prime +\bar X_{s}$
could be at the 10\% level and
easily observable at B Factories.

As noted some time ago, the low semileptonic branching ratio
(${\cal B}_{s.l.}$) and charm deficit ($n_C$) problems
could be explained by some hidden 
$B$ decay mode $\sim 10$--$15\%$ \cite{GH,Kagan}, 
such as $b\to sg$ with $g$ on-shell.
Two recent 
analyses \cite{KR,DISY} argue for
the existence of additional charmless $B$ decays at the 10\% level,
but differ in the interpretations.
Against the possibility of \cite{KR} $b\to sg \approx 10\%$  from new physics,
the authors of Ref. \cite{DISY} 
argue for nonperturbative effects from $b\to sc\bar c$ transitions.
Since the QCD penguin, known to be around 1\% \cite{HSS},
contains $b\to sc\bar c \to sg^* \to sq\bar q$ as a subprocess, 
one needs to invoke \cite{DISY} some intermediate $c\bar c$ state,
such as a $c\bar cg$ ``hybrid" meson. 
The ``hybrid" state, however, must satisfy the following:
1) sizable production fraction in $b\to sc\bar c$;
2) narrow width
to allow more time for $(c\bar c)_8 \to g^*$ annihilation;
3) 
decays via  $D\bar D + X$ or  $(c\bar c)_{\rm onia} +X$ 
are suppressed;
4) 
evasive so far in usual $e^+ e^-$ or $p\bar p$ annihilation studies.
Hence, 
though possible in principle,
the ``hybrid" (or any non-onia $c\bar c$ state) scenario is
no less exotic than the $b\to sg$ picture.

Several rare B decays
have just been reported \cite{CLEO,Miller} for the first time.
The penguin dominant $K\pi$ mode is at $1.5\times 10^{-5}$ level,
while the tree dominant $\pi \pi$ mode has yet to be seen.
This agrees with theory expectations \cite{Tseng}
if one takes into account the smallness of $V_{ub}$.
The $\omega K$, $\omega \pi$ modes are comparable to $K\pi$
and larger than expected.
No $\eta h^\pm$ events are seen, but 
$\eta^\prime K \simeq 7.1 \times 10^{-5}$ is quite sizable.
Together with $\phi K^* \sim K\pi$ \cite{Miller}, 
penguin effects are clearly rather strong.
Though interesting, these handful of modes do not yet
seriously challenge models of exclusive decay.
Perhaps more intriguing is the hint \cite{CLEO,Tom,Miller} of 
%
\begin{equation}
{\cal B}(B\to \eta^\prime + X_s) = (62 \pm 16 \pm 13) \times 10^{-5}
\mbox{\hskip 1.5cm \rm ($2.0 < p_{\eta^\prime} < 2.7$ GeV)}
\end{equation}
where $X_s \equiv K + n$($\leq 4$)$\pi$ 
with at most one $\pi^0$.
Some $\eta^\prime K$ events are captured,
but the $\eta^\prime K^*$ mode is conspicuously absent,
and most events are at larger $m_{X_s}$.
Though 
$\eta^\prime D^{\scriptstyle(*)}$, $\eta^\prime D^{**}$
backgrounds are yet to be fully ruled out \cite{Tom}, 
if large inclusive charmless {\it fast}
$\eta^\prime$ production becomes established soon,
it would be one of the most exciting piece of B physics ever.

As $\eta^\prime$ is mainly an SU$_{\rm F}(3)$ singlet, 
it is naturally related to gluons, motivating
Atwood and Soni \cite{AS} (AS) to
connect inclusive $\eta^\prime$ production to 
the QCD penguin via the gluon anomaly.
Denoting 
$\eta^\prime$-$g$-$g$ coupling as 
$H(q^2,k^2,m_{\eta^\prime}^2)\, \varepsilon_{\mu\nu\alpha\beta}\, 
q^\mu k^\nu\varepsilon^\alpha(q)\varepsilon^\beta(k)$,
they extract $H(0,0,m_{\eta^\prime}^2) \simeq 1.8$ GeV$^{-1}$ 
from $J/\psi\to \eta^\prime\gamma$ decay.
Assuming {\it constant} 
$H(q^2,0,m_{\eta^\prime}^2) \approx H(0,0,m_{\eta^\prime}^2)$,
they find that the standard $b\to s$ penguin could account for the 
$B\to \eta^\prime + X_s$ rate.
We wish to explore the $q^2$-dependence of the anomaly coupling, 
starting with the running of $\alpha_s$.

Let us give a more theoretical basis to the gluon anomaly coupling,
which concerns the $\eta^0$-$g$-$g$ effective vertex of 
the singlet field $\eta_0$.
In the chiral limit $m_q \rightarrow 0$
with $N_{\rm F} =3$ chiral quarks,
the singlet current has an anomaly,
$ 
\partial^\mu J_\mu^0 = (2N_{\rm F}\, {\alpha_s / 4\pi})\,
    {\rm tr}(G^{\mu\nu}\tilde{G}_{\mu\nu}),
$ 
which breaks the U$_{A}(1)$ symmetry,
and through the topological charge
$\left\langle 0\vert (2N_{\rm F} \alpha_s / 4\pi)
    {\rm tr}(G^{\mu\nu}\tilde{G}_{\mu\nu})
\vert \eta^\prime\right\rangle
$ etc., $m_\eta$ and $m_{\eta^\prime}$ 
are elevated by their ``gluon content".
Deriving from QCD the low energy effective theory of 
$\eta$ and $\eta^\prime$ mesons 
continues to be an active research field \cite{ag}.
The $\eta_0$-$g$-$g$ coupling can be formulated
{\it without assuming} PCAC, as the U$_{A}(1)$ symmetry is
already broken.
We find \cite{ag} the low energy effective coupling
$(N_{\rm F}\, \alpha_s / 4\pi)\, \theta\, 
    {\rm tr}(G^{\mu\nu}\tilde{G}_{\mu\nu})
$
which arises from the Wess-Zumino term,
where $\theta = \eta_0/\sqrt{N_{\rm F}}f_0$ is the collective
``chiral rotation". 
Both $\eta_0$ and the ``decay constant" $f_0$ are 
very complicated objects, and the connection of $\eta_0$ 
to physical mesons 
is highly nontrivial.
We saturate $\eta_0/f_0$ by 
$c_P\, \eta^\prime/f_{\eta^\prime} + s_P\,  \eta/f_\eta$,
where $s_P\equiv \sin\theta_P$ is the pseudoscalar mixing angle,
and sweep theoretical uncertainties such as 
form factor dependence into $f_{\eta^\prime}$.
We shall, however, assume constant 
$f_{\eta^\prime} \simeq f_\pi \simeq 131$ MeV \cite{AF}
in the following, which is still a rather strong assumption.

We arrive at the effective $\eta^\prime$-$g$-$g$ vertex,
\begin{equation}
-i\, a_g \ c_P\, \eta^\prime
\, \varepsilon_{\mu\nu\alpha\beta}\, 
\varepsilon^\mu(q) \varepsilon^\nu(k) q^\alpha k^\beta,
\end{equation}
where
$ 
a_g(\mu^2) = {\sqrt{N_{\rm F}}\, \alpha_s(\mu^2)/ \pi f_{\eta^\prime}}
$ 
is the effective gluon anomaly coupling and is nothing but
$H(q^2,k^2,m_{\eta^\prime}^2)$ of AS.
The explicit $\alpha_s$ factor suggests 
running coupling as commonly seen in QCD, 
a point which is ignored by AS.
As a check, we find $a_g(m_{\eta^\prime}^2) \simeq 1.9$ GeV$^{-1}$,
agreeing well with $H(0,0,m_{\eta^\prime}^2)$ found by AS,
which confirms the validity of Eq. (2).
With $q^2 \neq 0$ and $k^2 = 0$ but keeping $\eta^\prime$ on-shell,
however,
it is plausible that $\mu^2 = q^2$ for $q^2 > m_{\eta^\prime}^2$.

To compute the $b\to \eta^\prime sg$ rate, let us
define $v_i \equiv V_{is}^* V_{ib}$ and ignore $v_u $
(hence $v_t \cong V_{ts} \cong -V_{cb}$).
The loop  induced (see Fig. 1(a)) $b\to s$ current \cite{HSS} in SM is
\begin{equation}
{G_{\rm F} \over \sqrt{2}}{g_s \over 4\pi^2}\; v_t \;
\bar s t^a
\{\Delta F_1\; (q^2\gamma_\mu - q_\mu \not{\! q}) L
- F_2\; i\sigma_{\mu\nu} q^\nu m_b R \} b,
\end{equation}
where $ \Delta F_1 \equiv F_1^t - F_1^c \simeq 0.25
- ( -2/3\, \log(m_c^2/M_W^2)) \simeq -5$,
and $F_2 \cong F_2^t \simeq 0.2$.
Only the $F_2$ term contributes to on-shell $b\to sg$,
but since $F_2 \ll \vert \Delta F_1 \vert $,
$b\to sg^* \to sq\bar q$ dominates over $b\to sg$ \cite{HSS}.
Representing Fig. 1(a) as a box and Eq. (2) as a blob,
the $b\to \eta^\prime sg$ process \cite{AS} is shown in Fig. 1(b).
With $q^2 = (k+k^\prime)^2$ ($g^*$ mass)
and $m^2 = (p^\prime + k)^2$, 
the $sg\bar q$ system evolves into $X_s$ 
and $m^2 \equiv m_{X_s}^2$ is the 
{\it physical} recoil mass against the $\eta^\prime$ meson.
Because of the anomaly coupling, 
{\it a parton level calculation
gives us a handle on physical distributions}.
Defining $x\equiv m^2/m_b^2$, $y\equiv q^2/m_b^2$ and
$x^\prime\equiv m_{\eta^\prime}^2/m_b^2$,
we find
%
\begin{equation}
{d^{2}{\cal B}(b\to \eta^\prime sg) \over dxdy} \cong
0.2\left( {g_s(m_b) \over 4\pi^2}\right)^2 {a_g^2\, m_b^2 \over 4}
\left\{ \vert\Delta F_1\vert^2 c_0 
               + {\rm Re}(\Delta F_1 F_2^*) {c_1 \over y} 
               + \vert F_2\vert^2 {c_2 \over y^2}\right\},
\end{equation}
where 
$c_0 = (-2x^2y + (1-y)(y-x^\prime)(2x+y-x^\prime))/2$,
$c_1 = -(1-y)(y-x^\prime)^2$,
$c_2 = (2x^2y^2 - (1-y)(y-x^\prime)(2xy-y+x^\prime))/2$,
and the factor $0.2 \simeq V_{cb}^2 G_{\rm F}^2 m_b^5/192\pi^3 \Gamma_B$ 
comes from normalizing against 
${\cal B}_{s.l.}$ (see, e.g. ref. \cite{AS}).
We confirm the formulas of AS, but there are some 
subtle differences in defining $\Delta F_1$ and $F_2$,
to which we now turn.

AS adapt from leading order results from operator analysis.
They adopt the convention of Buras \cite{BBL} for the $c_8(\mu)$ coefficient
and absorb a factor of 1/2 into their definition of $F_2$.
In our notation, we find
$F_2(\mu) \simeq 0.286$ as compared to $F_2(m_t) \simeq 0.2$,
bringing
$ 
{\cal B}(b\to sg) 
$ 
from 0.1\% \cite{HSS} to 0.2\%.
This is agreeable since the dipole $O_8$ operator 
contains explicitly the gluon field.
The AS treatment of $F_1$ is more dubious. 
They identify
$4(c_4 + c_6)/g_s \equiv F_1^{\rm AS}(\mu) $ 
(which is our $ (g_s/4\pi^2)\Delta F_1$)
and find a value of $-0.168$ at LO \cite{BBL}.
In effect, they take the 
$(\bar s t^a\gamma_\mu Lb) (\bar q t^a \gamma_\mu q)$ part of
$c_4(\mu) O_4 + c_6(\mu) O_6$,
and replace $\bar q t^a \gamma_\mu q$ by a gluon.
This is, however, not appropriate
since the $c_4(\mu)$ and $c_6(\mu)$ coefficients
contain resummed leading logs:
the final $\bar s t^a\gamma_\mu Lb$ current does 
{\it not} simply couple to an effective gluon.
The correct approach is to 
insert $\eta^\prime$ in every step of the operator analysis, 
which is nontrivial and not yet done.

We will thus use the simple one loop results for $F_1$ and $F_2$ as
outlined earlier,
with $g_s = g_s(m_b)$ in Eq. (3). 
The operator analysis confirms that the correction is 
only of ${\cal O}(\alpha_s)$, 
but we now have the advantage of proceeding consistently 
from Fig. 1(a) to Fig. 1(b).
In addition, whereas the operator approach
usually stops at a set of effective operators at $\mu = m_b$,
our formalism automatically includes perturbative final state
rescattering effects such as $b\to sc\bar c \to sg^*$,
which is very useful when we turn to CP violating asymmetries.

Let us check against the results of AS numerically.
Using $m_b,\ m_s = 4.8,\ 0.15$ GeV,
$\alpha_s(m_b) \simeq 0.21$ 
and constant $a_g c_P \simeq 1.7$ GeV$^{-1}$,
we find that $(g_s/4\pi^2)\Delta F_1(\mu) = -0.168$ alone gives 
${\cal B}(b\to \eta^\prime sg) \approx 1.6\times 10^{-3}$,
not far from the AS result of $1.9\times 10^{-3}$. 
However, inclusion of
$F_2(\mu) = 0.286$ term leads to $\sim -20\%$ {\it reduction}
rather than the $\sim +50\%$ increase claimed by AS.
The formulas of AS in fact confirm our findings.
Note that the $F_2$ effect alone is negligible
but the interference effect is {\it destructive}
\cite{HSS}. 
The $d{\cal B}/dm$ plot in Fig. 3 \cite{AS} of AS
seems to be the dashed curve \cite{mg} for $d{\cal B}/dq$ in our Fig. 2(b).
The actual efficiency of an $m_{X_s}$ cut at 2.35 GeV (i.e. $p_{\eta^\prime}> 2$ GeV) 
for our $d{\cal B}/dm$ (Fig. 2(a)) 
is of order 1/2, and is {\it not sensitive} to Fermi motion of the $b$ quark.
The $b\to \eta^\prime sg$ rate 
for pure $\vert F_2\vert \cong 2$ 
(i.e. $b\to sg \sim 10\%$ from new physics) 
is slightly lower than the SM result.
But if it interferes with $\Delta F_1^{\rm SM}$ constructively, 
the resulting ${\cal B}(b\to \eta^\prime sg) \simeq 0.8\%$ would be
way too large.

However, as argued earlier, 
the anomaly coupling $a_g \propto \alpha_s$ could be running. 
Since $(m_b - m_s)^2 \geq q^2 \geq m_{\eta^\prime}^2 $, 
the likely scale would be the $q^2$ of the virtual gluon.
Using two-loop running $\alpha_s(q^2)$ in $a_g$, 
${\cal B}(b\to \eta^\prime sg)$ drops by a factor of more than 3.
This is because $\alpha_s(m_b^2)/\alpha_s(m_{\eta^\prime}^2) \sim 1/2$,
and the derivative coupling nature of the anomaly 
favors large $q^2$ and $m^2$, as seen in Fig. 2.
The anomaly is thus {\it uniquely suited
for generating fast $\eta^\prime$ mesons}.
The SM effect alone now drops to $\sim 0.43\times 10^{-3}$, and, 
even without applying the $m_{X_s}$ cut,  it falls short of Eq. (1).
Thus, Fig. 1(b) with running $\alpha_s$ in the anomaly coupling 
suggests that new physics is needed 
to account for the observed $B\to \eta^\prime + X_s$ rate\cite{etac}.

It is possible to enhance the chromo-dipole $bsg$ coupling
by new physics at the TeV scale (such as supersymmetry or techniscalars),
without jeopardizing the electro-dipole $bs\gamma$ coupling
\cite{Kagan,CGG}.
Explicit examples \cite{CGG} with gluino loops 
favor large $F_2^{\rm New}$ with sign {\it opposite} to SM, 
which is just what is needed (see below).
The chromo-dipole term may be linked to quark mass generation
since both involve chirality flip \cite{Kagan}.
This TeV scale connection and the appearance of the $b_R$ field
provide an exciting impetus to the problem, namely CP violation
\cite{KaganCP}.
One is now {\it sensitive to CP violating phases which are in principle 
different from the standard CKM phase}.
Note that within SM, the small effect of $v_u\neq 0$
leads to CP violating asymmetry $<1\%$ in $b\to \eta^\prime sg$,
much like other inclusive $b\to s$ decays \cite{CPT}.

The CP violation effect is precisely rooted in 
$\Delta F_1^{\rm SM}$-$F_2^{\rm New}$ interference.
Parametrizing the new physics term as
$F_2 \equiv F_2^{\rm SM} + F_2^{\rm New} \simeq -2\, e^{i\sigma}$
with $v_t$ taken as real,
the required absorptive part comes from \cite{CPT,BSS}
$c\bar c$ rescattering in $\Delta F_1$ (see the cut in Fig. 1(a)). 
This is facilitated by the peaking of $d{\cal B}/dq$ at 
$q^2 
{\ \lower-1.2pt\vbox{\hbox{\rlap{$>$}\lower5pt\vbox{\hbox{$\sim$}}}}\ }
 (2m_c)^2$.
The absorptive part is incorporated 
by making the change \cite{CPT}
$\Delta F_1 \to \Delta F_1+ 4\Pi(q^2/m_c^2)$,
where $\Pi$ is the familiar one-loop vacuum polarization from QED.
For $\bar b\to \eta^\prime \bar sg$ one simply
replaces $F_2^*$ by $F_2$ in Eq. (4).
We thus easily arrive at the average branching ratio 
${\cal B}_{\rm av.}$ and asymmetry 
$a_{\rm CP} = ({\cal B} - \bar {\cal B})/ ({\cal B} + \bar {\cal B})$,
and the results are given in Table 1.
The asymmetry is generally larger for $\cos\sigma < 0$
(except vanishing as $\sigma \to \pi$) because of
destructive interference, but ${\cal B}_{\rm av.}$ 
often becomes too small in this region.
To visualize the effect, we give in Fig. 3 
the Dalitz plot (in $q$ and $m$) and differential rates for
both  $b\to \eta^\prime sg$ and $\bar b\to \eta^\prime \bar sg$,
for $\sigma = \pi/2$.
The more visible difference in $d{\cal B}/dq$
is not observable. 
However, since the shape for $d{\cal B}/dm$ is largely
unchanged, a 10\% difference in rate below the 
$m_{X_s}$ cut of 2.35 GeV should be readily visible,
at CLEO and at proposed B Factories 
that would start operation in 1999.
This is a {\it direct} CP violation effect independent of
$B^0$-$\overline B^0$ mixing, and can be seen in both charged and
neutral $B$ decays,
{\it in a mode which has already been observed}.

Some remarks are in order.
First, 
$b\to sg\sim 10\%$ alone leads to
$b\to \eta^\prime sg$ only at $\sim 0.5\times 10^{-3}$,
comparable to the standard $b\to sg^*$ penguin 
which starts from 1\%.
This is because $\Delta F_1^{\rm SM} \sim 5$ is still larger than
$\vert F_2^{\rm New}\vert\sim 2$.
Second, 
in our numerical study, we took $m_g \sim 0.5$ MeV in phase space 
to remove soft gluons \cite{mg}.
If one assumes the $sg\bar q$ system with a soft gluon (see Fig. 2(a)) 
is swept into the $K$ meson, the removed 4-5\% matches 
the observed exclusive $\eta^\prime K$ rate.
Third, 
the 
$B\to \eta + X_S$ rate should be smaller by $\tan^2\theta_P \sim 0.1$ in rate.
However, fast $\eta$ from the $B\to \eta^\prime + X_S$, 
$\eta^\prime \to \eta\pi\pi$ cascade may be the source of the
little ``bump" at high $p_\eta$ in 
fully inclusive $B\to \eta + X$ spectrum
\cite{eta}.
Four,
the ``hybrid $c\bar c g$" mechanism of
Ref. \cite{DISY} might also work, 
since the effective $b\to sc\bar c\to sg^*$ penguin is
much larger than in perturbative applications of SM \cite{etac}.
However, SM mechanisms alone would never bring about 
CP asymmetries beyond 1\% in these modes \cite{CPT}.
Thus, {\it the large CP asymmetries discussed here could
serve as a unique signature for the presence of 
new physics from dipole $b\to sg$ transition} \cite{KaganCP}.
Five, 
the existence of large $b\to sg$ and associated CP asymmetries
would have implications on the responsible new physics at TeV scale.
For example, the lightest squark could still be around 100 GeV
and gluino mass of order 200 GeV. 
Existing bounds are evaded by large nondegeneracies in squark masses,
but such masses can certainly be probed at the Tevatron.
However, there are solutions where $m_{\tilde q}$ and $m_{\tilde g}$ 
are much higher \cite{Kagan,CGG}.
Finally, 
we stress that the anomaly induced $b\to \eta^\prime sg$ of Fig. 1(b)
is a new diagram {\it in addition} to the usual $b\to sg^*\to sq\bar q$,
and shows very different $q^2$ and $m^2$ dependence.
It is worthwhile to pursue effects of the gluon anomaly
in more conventional processes, in particular, to measure
$\eta^\prime$-$g$-$g$ form factor effects.
An example would be $e^+e^-\to \eta^\prime + q\bar q g$,
to see if gluon fragmentation into $\eta^\prime$ differs from, say, into pions.
The study of these processes would be reported elsewhere.

In summary, applying the $\eta^\prime$-$g$-$g$ anomaly
with running $\alpha_s$, we find that 
${\cal B}(b\to \eta^\prime_{\it fast} + X_s) \sim 0.6\times 10^{-3}$ 
perhaps cannot be sustained by the standard $b\to sg^*$ penguin alone,
but calls for new physics from $b\to sg$ at 10\% level,
which would also help alleviate the ${\cal B}_{s.l.}$ and $n_C$ problems. 
{\it Direct} CP violating rate asymmetries
could then be as large as 10\% and easily observable,
perhaps even before the advent of asymmetric B Factories.

\acknowledgments
We thank T. E. Browder, A. L. Kagan, Y. C. Kao, G. Veneziano
and H. Yamamoto  for discussions,
and H. C. Huang for help in graphics.
This work is supported in part by 
grants
NSC 86-2112-M-002-019 and NSC 86-2112-M-002-010-Y
of the Republic of China.

\vskip0.9cm
\noindent {\bf Note Added.}
After our work was submitted, new experimental \cite{Miller}
and theoretical results \cite{AS2,KP,Fritzsch} have appeared.
We thank A. L. Kagan for correcting us for the
{\it destructive} nature of the $\Delta F_1 F_2$ term in SM
\cite{HSS,KP}, which strengthens the need for $F_2^{\rm New}$.
Our work has aroused interest in the off-shell (form factor) behavior of
the $g^*$-$g$-$\eta^\prime$ vertex.
Ref. \cite{KP} argues for $m^2_{\eta^\prime}/(m^2_{\eta^\prime} - q^2)$ 
form factor suppression in analogy to 
the quark triangle loop for $\gamma^*$-$\gamma$-$\pi^0$.
However, while asymptotically $1/q^2$ suppression must set in,
the gluon anomaly differs from the QED case
in the gluon self-couplings.
Indeed, Refs. \cite{AS2} and \cite{Fritzsch} stress that 
nonperturbative effects could make the $g^*$-$g$-$\eta^\prime$
vertex unpredictable (Ref. \cite{AS2} extends our 
new physics CP violation effect to channels beyond $\eta^\prime$).
The $\sqrt{q^2} \sim 2$--$4$ GeV range of interest coincides with 
the glueball mass range, which might well delay the onset of
form factor suppression, e.g.  in the form of $m^2_{G}/(m^2_{G} \pm q^2)$
where $m_G$ is the relevant glueball mass.
On the other hand, the anomaly coupling is fixed at 
the (low energy) $m_{\eta^\prime}$ scale,
and it seems unlikely that one would hit a resonance pole in $q^2$.
In any case, the $g^*$-$g$-$\eta^\prime$ form factor at intermediate $q^2$
is an extremely interesting subject in QCD itself, 
and has yet to be studied.
However, even with nonperturbative $g$-$g$ binding effects,
running $\alpha_s$ should still be taken into account
since it enters multiplicatively.
Our criticism of AS is that, 
taking  $H(q^2,0,m_{\eta^\prime}^2) = H(0,0,m_{\eta^\prime}^2)
 = a_g(m_{\eta^\prime}^2)$ to be constant and equal to
the largest possible coupling, 
they likely overestimate the SM effect.
To the least, our work can be viewed as an illustration 
of how a consistent picture of 
semi-inclusive $B\to \eta^\prime_{\rm fast} + X_s$ and the
${\cal B}_{s.l.}$ and $n_C$ problems together suggest
a common $b\to sg \sim 10\%$ new physics solution,
and the dramatic consequence of 
$\sim 10\%$ {\it inclusive direct} CP asymmetries
that might follow.


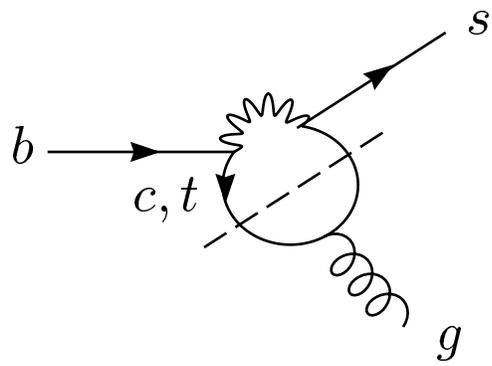
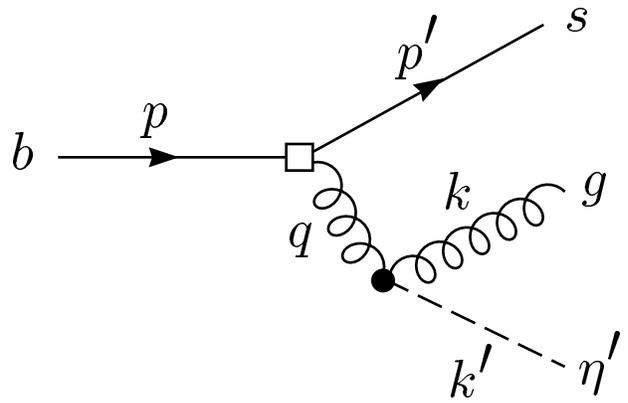
\begin{figure}
\caption{
(a) Sample diagram for loop induced $b\to sg^*$
with possible $c\bar c$ cut;
(b) $b\to \eta^\prime sg$ transition via effective $b$-$s$-$g$ coupling
(possibly from new physics)
and $\eta^\prime$-$g$-$g$ anomaly vertex.
}
\label{fig1}
\end{figure}

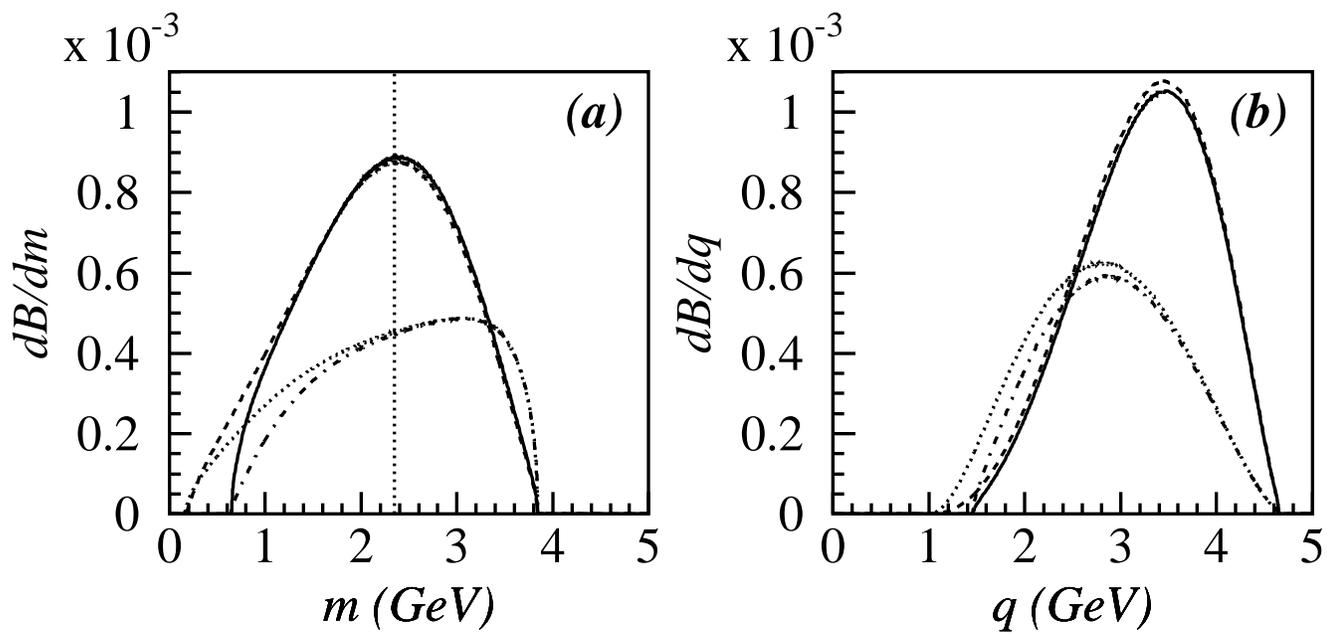
\begin{figure}
\caption{
(a) $d{\cal B}/dm\equiv d{\cal B}/dm_{X_s}$ and (b) $d{\cal B}/dq$
for SM penguin induced $b\to \eta^\prime sg$
(dashed, solid: cut of $m_g = 0,\ 0.5$ GeV).
The purely dipole (dotdash) effect with $\vert F_2\vert\approx 2$ 
is also given.
The vertical dotted line indicates the $m_{X_s} = 2.35$ GeV cut.
}
\label{fig2}
\end{figure}

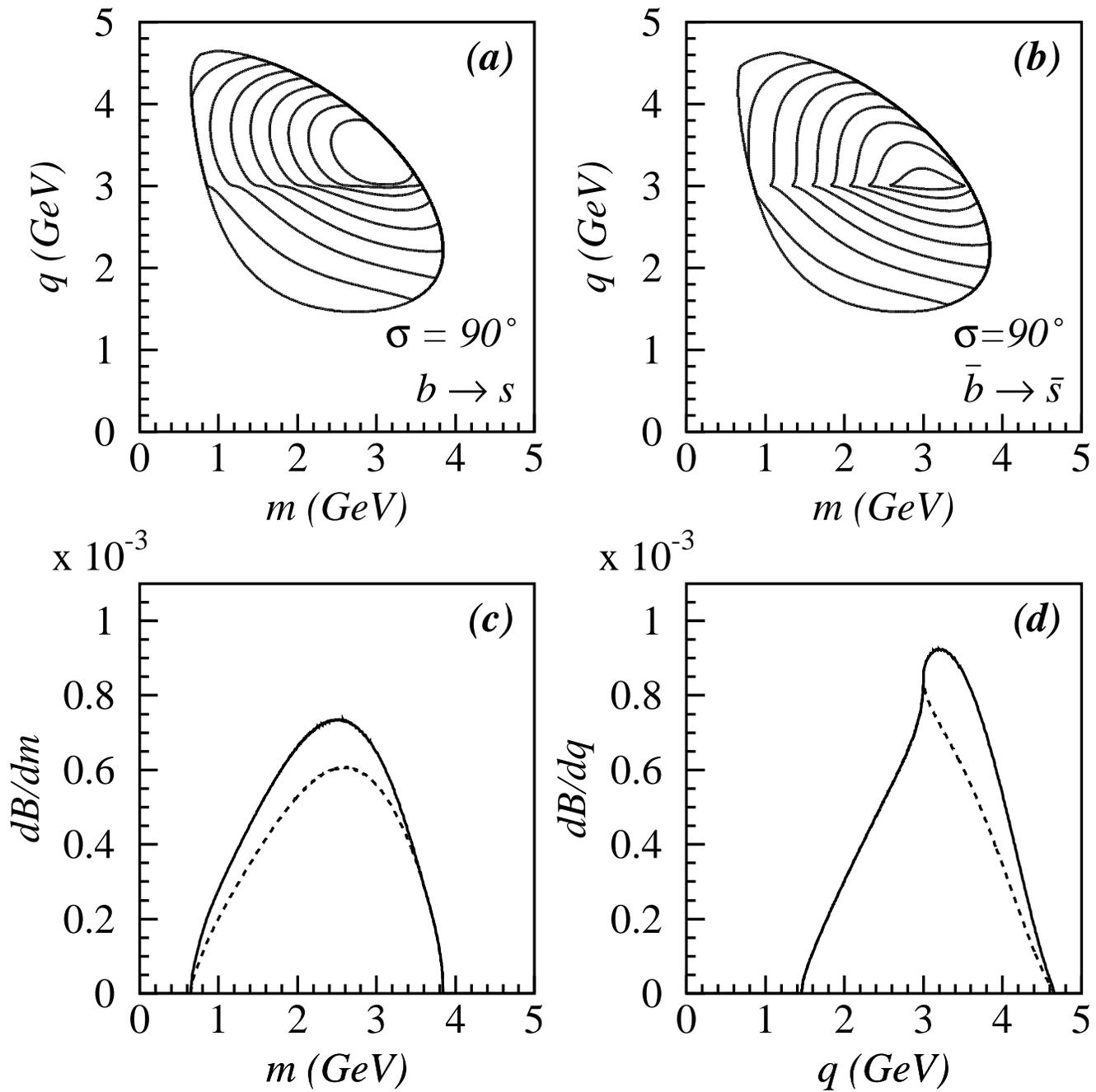
\begin{figure}
\caption{
(a), (b) Dalitz plot and (c) $d{\cal B}/dm$ (d)  $d{\cal B}/dq$ for
$b\to \eta^\prime sg$ (solid) vs. $\bar b\to \eta^\prime \bar sg$
(dashed), for $\sigma = \pi/2$.
}
\label{fig3}
\end{figure}

\begin{table}
\caption[] {
${\cal B}_{\rm av.}$ and $a_{\rm CP}$ for 
$b\to \eta^\prime sg$ vs. $\bar b\to \eta^\prime \bar sg$ transitions,
with $F_2^{\rm New} = -2 e^{i\sigma}$.
The latter alone gives branching ratio 
$\simeq 0.45\times 10^{-3}$,
comparable to SM effect without $c\bar c$ rescattering.
}
\begin{tabular}{ccccccccd}
  $\sigma = $ & $0$ & $30^\circ$ & $60^\circ$
                & $90^\circ$ & $120^\circ$ & $150^\circ$ & $180^\circ$ \\ \hline
  ${\cal B}_{\rm av.}$ ($\times 10^{-3}$)
     &  2.5 & 2.4 & 2.0 & 1.5 & 0.91 & 0.52 & 0.4 \\ \hline
  $a_{\rm CP}$ (\%)
     & 0.0  & 2.9 & 6.0 & 9.5 & 13.1 & 13.3 & 0.0 \\
\end{tabular}
\label{table1}
\end{table}


\begin{references}
\bibitem{mg} A cut of $m_g = 0.5$ GeV (in phase space) 
removes soft gluons which are ill-defined.
The reduction by 4-5\% is small since there are no IR singularities.
\end{references}
\end{document}